# Impact Ionization and Carrier Multiplication in Graphene

Luca Pirro[1,2], Anuj Girdhar[1,3], Yusuf Leblebici[2] and Jean-Pierre Leburton[1,3,4]

[1]Beckman Institute, University of Illinois at Urbana-Champaign, Urbana, Illinois, 61801, USA
[2] Microelectronic System Laboratory, EPFL, Lausanne, CH 1015, Switzerland
[3]Department of Physics and
[4]Department of Electrical and Computer Engineering, University of Illinois at Urbana-Champaign, Urbana, Illinois, 61801, USA.

**Abstract**

We develop a model for carrier generation by impact ionization in graphene, which shows that this effect is non-negligible because of the vanishing energy gap, even for carrier transport in moderate electric fields. Our theory is applied to graphene field effect transistors for which we parametrize the carrier generation rate obtained previously with the Boltzmann formalism [A. Girdhar and J. Leburton, Appl. Phys. Lett. **99**, 229903 (2011)] to include it in a self-consistent scheme and compute the transistor I-V characteristics. Our model shows that the drain current exhibits an "up-kick" at high drain biases, which is consistent with recent experimental data. We also show that carrier generation affects the electric field distribution along the transistor channel, which in turn reduces the carrier velocity.

**I. Introduction**

In recent years, graphene has emerged as a new electronic material with unusual physical properties, due to its two-dimensional (2D) nature and its band structure, where the carrier energy is linear in momentum and the gap separating conduction and valence bands is reduced to the single Dirac point [1]. For this reason, a large amount of work has been devoted to explore new physical effects and exploiting them in various technological applications [2-7]. In this context, graphene's 2D nature and high carrier velocity are well suited for high speed and high performance electronics, for which the interaction between charge carriers and static and dynamic lattice defects has been studied well [8]. By contrast, less attention has been paid to interband interaction amongst carriers [9]. In conventional semiconductors, it is well known that this kind of inter-carrier interaction is characterized by an energy threshold of the order of the energy gap, which restricts the energy exchange amongst carriers to the most energetic ones and thus becomes significant in high electric fields [10]. In gapless graphene such a condition is not fulfilled.

Very recently, electron-hole generation rates caused by interband carrier-carrier interaction in the presence of electric fields have been obtained (AG and JPL), and this confirm the absence of energy threshold for impact ionization [11]. Moreover, it was shown that the generation rate is quasi-quadratic in the electric field at constant carrier temperature and strongly decreases with the carrier concentration as it reduces the density of final scattering states for both particles. Therefore, it becomes evident that any analysis of the transport characteristics of graphene without the consideration of impact ionization and its consequence on the carrier concentration is incomplete.

In this paper, we provide a physical model that takes into account impact ionization in non-linear transport in graphene and graphene-based devices, specifically field-effect transistors that are mostly utilized to extract transport parameters as a function of carrier concentrations [12].

## II. Impact Ionization-Limited Transport Model

Let us consider a graphene sheet placed in an electric field $F$ in the $x$-direction. At steady state, and in the presence of impact ionization, both electron and hole current densities $J_n$ and $J_p$ satisfy the 1D continuity equations

$$\frac{dJ_n}{dx} + eG(x) = 0 \quad (1.a) \qquad \text{and} \qquad \frac{dJ_p}{dx} - eG(x) = 0 \quad (1.b)$$

where $G(x)$ is the net electron-hole pair (EHP) generation. Combining these two equations yields

$$\frac{d(J_p + J_n)}{dx} = 0 \quad (2.a) \qquad \text{and} \qquad \frac{d(J_p - J_n)}{dx} - 2eG(x) = 0 \quad (2.b)$$

where eq.2.a expresses the total current conservation, and $2G(x)=U$ is the particle generation rate given by [11]

$$U = -\frac{8}{A} \sum_{\substack{k_1,k_{1'} \\ k_2,k_{2'}}} S(k_1,k_{1'};k_2,k_{2'}) \left\{ \begin{array}{l} [f(k_{1'})][f(k_{2'})][1-f(k_1)][1-f(k_2)] \\ -[f(k_1)][f(k_2)][1-f(k_{1'})][1-f(k_{2'})] \end{array} \right\} \delta_{k_1+k_2,k_{1'}+k_{2'}} \quad (3)$$

where

$$S(k_1,k_{1'};k_2,k_{2'}) = (2\pi/\hbar) \qquad E_1 + E_2 - E_{1'} - E_{2'}). \quad (4)$$

Here $M = e^2/(2x_{eff}x_0 q\varepsilon(q))$ where $k_{eff}$ is the effective dielectric constant of a single layer graphene on a insulating substrate, and $\varepsilon(q)$ is the static screening dielectric function of the $q=k_1-k_{1'}$ wave vector. The factor 8 accounts for spin and valley degeneracies as well as two distinct particles, electron and hole. $k_1$ and $k_2$ are the initial wave vectors of electrons in the conduction (C) band and valence (V) band, respectively, and $k_{1'}$ and $k_{2'}$ are the final wave vectors in the conduction band (CVCC process) (Fig. 1). A similar equation also describes the VCVV process. $S(k_1, k_1'; k_2, k_2')$ is the transition probability per unit time for interband carrier-carrier scattering, $f(k)$ is the carrier distribution function assumed to be a displaced Fermi-like distribution with an electronic temperature [11-13], and A is the sample area. The reduction of the 8-uple summation in eq.3 to a quadruple integral to obtain the generation rate numerically is demonstrated in Appendix A. Fig. 2 illustrates the quasi-quadratic variation of the particle generation rate $U$ with electric fields for different carrier concentrations and electronic temperatures. In particular, it shows that $U$-rates increase (decrease) with temperature (carrier concentration) as the latter increases (decreases) the density of available final states for scattered electrons [11].

The generation rate obtained numerically from eq.A.13 is cumbersome for the integration of eqs.2. For this reason, we propose the following expression

$$U = gF^\alpha \exp[-(n/n_0)] \qquad (5)$$

where $g$, $n_0$ and $\alpha \sim 1.66-1.75$ are fitted from Figs. 2 data. The $\alpha$-value is slightly smaller than 2 (Fig. 3.a), because $U$ deviates from the strictly quadratic dependence on the electric field at high fields [11] and slightly increases with carrier concentration. For the sake of simplicity we will use $\alpha=1.7$. Fig. 3.b shows the exponential variation of $U$ with carrier concentration for different temperatures, where the $n_0$-parameter is a linear function of the electronic temperature (Fig. 3.c). It was also found that a variation $g \propto T^{3/2}$ on the electronic temperature constitutes a good approximation. Finally, one can relate the electronic temperature ($T_e$) dependence on the electric fields by the usual quadratic expression [14]

$$T_e = T_L[1 + (\frac{F}{F_{CT}})^2], \qquad (6)$$

where $T_L$ is the lattice temperature, and $F_{CT}$ is a critical field for the onset of hot carrier effects, which depends on various scattering mechanisms.

**III. Impact Ionization in Graphene Field Effect Transistor**

In order to compare our carrier generation model with available experimental data on high field transport, we consider the standard configuration of a graphene field-effect transistors (G-FET) shown in Fig. 4 for a n-channel. In the charge-control model, one can write quite generally,

$$Q_+ - Q_- = \pm C_{ox}(V_{GT} - V(x)) \tag{7}$$

for the ambipolar nature of the graphene channel charge, where $Q_+$ and $Q_-$ are the hole and electron charges respectively, $C_{ox}$ is the oxide capacitance, and $V_{GT} = V_G - V_T$, where $V_G$ is the gate bias and $V_T$ is the gate voltage at minimum conductance. The $\pm$ sign is for electron (+) and hole (-) channels, so that $V_{GT}$ is positive and negative, respectively, and $V(x)$ is the potential drop along the channel (source at $x=0$). We define the hole/electron current as $I_{p,n} = \pm Q_{+,-} W v_{p,n}(F)$ [15], where $W$ is the graphene channel width, and $v_{p,n}(F)$ is the carrier velocity given by [16],

$$v(F) = \pm \frac{\mu_0 F}{1 + \frac{F}{F_C}} \tag{8}$$

where the $\pm$ sign is for holes (+) and electrons (-), respectively, and $F$ is the electric field. The parameters $\mu_0$ and $F_C$ are the low field mobility and the critical field for the onset of non-linearity due to high energy carrier scattering such as by optic phonons [17]. Both are assumed equal for electrons and holes given the symmetrical band structure of graphene. We also assume $F_C$ and $F_{CT}$ may be different as current non-linearity and electronic temperature onset may have different origins; the former is related to the carrier momentum relaxation whereas the latter is related to the energy relaxation [14]. Neglecting any diffusion processes, and integrating eq.(2.b), we get

$$I_p^{drift} - I_n^{drift} - eW \int_x^L U(x)dx = I. \tag{9}$$

By using eq.7 and the hole current definition,

$$\frac{C_{0x} W \mu_0 (V_{GT} - V) F}{1 + \frac{F}{F_C}} + 2eW \int_0^x G(x) dx = I \quad (10)$$

where $I$ is the total current. By integrating eq.10 along the channel length ($L$), as usually done in the charge control model of conventional MOS device [15], we obtain the expression for the total current in the transistor is (see Appendix B)

$$I = \frac{C_{ox} W \mu_0 \left[ V_{GT} V_{DS} - V_{DS}^2/2 \right]}{L(1 + \frac{V_{DS}}{V_C})} + \frac{2eW}{1 + \frac{V_{DS}}{V_C}} \int_0^L \left( 1 - \frac{x'}{L} + \frac{V_{DS} - V(x)}{V_C} \right) G(x') dx' \quad (11)$$

with $V_C = F_C L$. Additionally, the expression of the electric field as a function of distance along the channel length [18] can be derived from eq.10 and is given by (see Appendix C)

$$F(x) = -\left\{ \frac{\frac{2eG(x)}{F_C C_{ox} \mu_0} - \frac{2eG(x)}{F_C C_{ox} \mu_0} \left[ V_{GT} - V_i - V(x) + \frac{2e}{F_C C_{ox} \mu_0} \int_0^x G(x') dx' \right] + \frac{2e}{C_{ox} \mu_0} \int_0^x G(x') dx' - F_C V_i}{\sqrt{\left[ V_{GT} - V_i + \frac{2e}{F_C C_{ox} \mu_0} \int_0^x G(x') dx' \right]^2 - 2 F_C V_i x + \frac{4e}{C_{ox} \mu_0} \int_0^x \left[ x - x' + \frac{V(x)}{F_C} \right] G(x') dx'}} \right\} \quad (12)$$

where $V_i = \frac{I}{W \mu_0 F_C C_{ox}}$, and which is an implicit function of the potential $V(x)$, so that eqs.11-12 are solved by iteration for a particular current $I$ value.

**Electric field and potential extrapolation beyond pinch-off**
Eqs.7, 10-12 are obtained under the gradual channel approximation within the charge control model (CCM) that ignores potential and electric field spatial variations beyond the channel pinch-off once current saturation is achieved [8]. As carrier generation by impact ionization mostly occurs in the high field region close to the drain, the distribution of the electric field and potential in that region should be assessed. For this purpose one can fit $F$ along the channel given by the actual CCM with a quadratic expression

$$F(x) = F_0 + ax + bx^2 \qquad (13)$$

where $F_0$ is the field at the source and $a$ and $b$ coefficients are function of $V_{DS}$ and $V_{GT}$ as shown in Fig. 7. Eq.13 is a good approximation on the source side of the transistor, but it underestimates the field on the drain side (see section VI). By integration one gets the corresponding expression for the electric potential

$$V(x) = -(F_0 x + \frac{ax^2}{2} + \frac{bx^3}{3}) \qquad (14)$$

with $V(L) = V_{DS}$ and $V(0)=0$.

One can use self-similarity with expressions (13-14) for two channels of different lengths $L$ and $L'$, so that electric potentials and fields are related by the following equations

$$\begin{aligned} V(L;a,b) &= V(L';a',b') \\ F(L;a,b) &= F(L';a',b'), \end{aligned} \qquad (15)$$

yielding

$$\begin{pmatrix} a' \\ b' \end{pmatrix} = \frac{6}{L'^3} \begin{pmatrix} L'L\left(\frac{L}{2}-\frac{L'}{3}\right) & L'L^2\left(\frac{L-L'}{3}\right) \\ L\left(\frac{L'-L}{2}\right) & L^2\left(\frac{L'}{2}-\frac{L}{3}\right) \end{pmatrix} \begin{pmatrix} a \\ b \end{pmatrix} + \frac{6F_0}{L'^2} \begin{pmatrix} L-L' \\ \frac{L'-L}{L'} \end{pmatrix}. \qquad (16)$$

With this approach one can then extrapolate the electric field and potential spatial profiles beyond pinch-off by asserting for $V_{DS}(L')>V_{DSAT}(L')$, $V_{DS}(L')=V_{DSAT}(L)$ if $L>L'$ and, similarly, for $F(L)$ and $F(L')$. From the field distribution, the carrier concentration along the channel

$$n(x) = \frac{I}{Wev(F(x))} \quad (17)$$

is readily obtained by the usual definition of current.

## IV: Computational approach

Because of the interdependence amongst generation rate ($G$) (eq.5), carrier concentration (eq.17), and potential and electric field profiles (eq.12), we use a self-consistent loop to compute the $I$-$V$ characteristics of the G-FET (Fig. 6) in this scheme. For each $V_{DS}$ value, we initialize the loop by computing the current (eq.11) in the absence of carrier generation to obtain the electric field and potential (eq.C.2) along the channel, which we then fit with eqs.13-14 to get the values of $a$- and $b$-coefficients. We then use eqs.15 to find $L'$, and then we re-calculate $a'$ and $b'$ beyond pinch-off. From these new sets of values we compute the initial $G$-profile, which we use to re-calculate the electric fields (eq.15), carrier concentration (eq.17) and total current. The iteration process is repeated until convergence of $I$ and $n$.

## V. Results

In order to illustrate the effects of carrier multiplication in graphene under high fields, we consider a G-FET with the following parameters: $W=2\mu m$, $L=1\mu m$, $\mu_0=2000 cm^2/(Vs)$, $C_{ox}=500 nC/cm^2$, $F_C=15 kV/cm$ and $V_T=0.5V$ [18]. We perform the simulation for an n-channel (positive top-gate voltage), and positive drain source bias, but the model is valid for a p-channel as well as long as we invert the signs of the biases.

Fig. 7.a illustrates the effect of carrier multiplication on the I-V characteristics of the G-FET for three values of the gate voltage, which shows an "up-kick" in the current at high drain bias beyond pinch-off. For the three gate biases, the corresponding saturation voltages are $V_{DSAT}=$ 0.79V, 1.11V and 1.38V, successively. Here we also assume a constant critical field $F_{CT}$

=25kV/cm for the onset of electronic temperature (eq.6) for all gate and drain biases. From the figure one can also see that the higher the gate bias, the weaker the effect as the value of the excess current slightly decreases compared to the saturation value. In Fig. 7.b we display the values of the electric field at the drain side as a function of drain bias for the three different gate biases. Interestingly, we note that the field continues to increase and remains finite even beyond pinch-off unlike what was predicted in a conventional CCM, where it remains constant at the saturation value. Furthermore, the lower the gate bias, the higher the field, which partially explains the decrease of the current "up-kick" in Fig. 7.a. However, there is also an influence on the carrier concentrations, which is a strong condition for the onset of carrier generation (see e.g. Fig. 2 and eq.5).

Fig. 8.a shows the net generation rate along the channel for different drain biases at a fixed gate voltage $V_{GT}$=1V, which increases with the drain source voltage, but also shows a quasi exponential increase away from the source as the field and electronic temperature (right axis) increase toward the drain. One notices however a tempering of the generation rate toward the drain at high drain source bias, which is due to the increase in the carrier concentration that limits the rate according to eq.5. In Fig. 8.b, we show the effect of the critical field for the onset of electron temperature $F_{CT}$ on the generation rate. As expected, the rate decreases with increasing $F_{CT}$ as the electronic temperature decreases, which weakens the impact ionization process and carrier multiplication. It is the most pronounced on the drain side.

**Comparison with experiment**

In their seminal 2008 paper, Meric *et al.* [12] reported features similar to those shown on Fig. 7 in the I-V characteristics of their double gate G-FET. In order to compare our model with their experimental data, we modify our approach to account for source and drain series resistances, and consider a p-channel. For this purpose we follow the approach developed by Scott and Leburton [18], where the first part of eq.11 becomes (for holes)

$$\frac{C_{ox}W\mu_0\left[V_{GT}V_{DS}-V_{DS}^2/2\right]}{L\left(1+\frac{V_{DS}}{V_C}\right)} \rightarrow \frac{V_{DS}-V_C+IR_S+\sqrt{(V_{DS}-V_C+IR_S)^2-4IR_SV_{DS}}}{4R_S} \quad (18)$$

where $I$ is the current in the system and $R_S$ the series resistance. For p-channel, the current is negative, and the sign changes in front of the integral term (eqs.11-12). From eq.18 it is clear that there will be a new term in the square root of eq.12,

$$[V_{GT}-V_i]^2 \rightarrow [V_{GT}-V_i-IR_s]^2 \quad (19)$$

We also add the ohmic drop $IR_S$ from the source in the potential (eq.14).

Figs. 9 compare our model calculations (solid lines) with the experimental data (stars) for two back gate voltages. For both cases we use the same values of the critical field for onset of velocity non-linearities as in Scott *et al.* [18]. In Fig. 9.a. ($V_{gback}$=-40V), the two current curves for low top gate biases ($V_{gtop}$=-0.3V & -0.8V) are fitted quite well with $F_{CT}$=22kV/cm, for which the effect of carrier generation by impact ionization is weak. For the highest top gate bias ($V_{gtop}$=0V), the best fit is obtained with a lower critical field, $F_{CT}$=16kV/cm, which is understandable since the hole concentration is smaller, which enhances carrier generation. There is a clear "up-kick" due to the generation rate at high source drain (negative drain) bias in good agreement with the experimental data. Fig. 9.b displays the comparison between model and experiment for $V_{gback}$=40V and three different top gate biases. We obtain a very good agreement with $F_{CT}$=22kV/cm for the lowest top gate voltage ($V_{gtop}$=-1.3V), for which carrier multiplication is weak owing to the high hole concentration. The agreement is less evident for the two lower current curves, where the best fit is obtained for $F_{CT}$=15kV/cm. Here again the effect of carrier generation is stronger for the highest top gate bias, because the hole concentration in the channel is the lowest while the discrepancy is also the largest. Our analysis shows that a change in the series resistance (dashed curve), not the low field mobility nor the critical field $F_C$ in eq.8, results in a better agreement between theory and experiment at high source drain bias but overestimates the conductance at low source drain bias in this case [18].

Fig. 10 displays the hole concentration on the drain side as a function of the source-drain bias for three top gate biases. As the latter increases, the former decreases monotonously to reach its minimum value at the onset of saturation, where, according to conventional CCM and in absence of carrier generation (dashed lines), it remains constant beyond pinch-off. It increases again at higher source drain bias due to carrier generation. We also plot the carrier drift velocity along the channel at saturation onset for the three top gate biases, where one also can see that lower velocity values are achieved at the drain side, where the velocity is saturating as well.

## VI. Conclusions

Because of the vanishing energy gap, carrier multiplication by impact ionization takes place in graphene without a carrier energy threshold, even in moderate electric fields ($F \geq 20-30 kV/cm$) which affects the transport characteristics at low carrier concentrations. Our theory based on a parametrization of the carrier generation rate within an extended charge control model shows that this effect is observable in graphene field effect transistors as a smooth "up-kick" in the current characteristics mostly at low gate voltage and moderate drain bias. A higher gate voltage lends to a higher drain bias for current "up-kick". We also showed that this effect is self-limited as it reduces the electric field, and consequently, the carrier velocity variation in the region along the channel where it takes places as a result of current conservation. One of the main assumptions of our model is the parametrization of the electric field (eq.13) as a function of distance from the source along the channel, which reduces the field on the drain side compared to the CCM (Fig. 7). However, we do not believe this effect alters our conclusions as these discrepancies occur over a short distance, while it is well known that the CCM leads to unphysical large fields on the drain side beyond current saturation [8]. Moreover, as mentioned previously, carrier generation itself softens the increase of the field on the drain side, which tends to validate eq.13.

## V. Acknowledgment

L. Pirro would like to acknowledge the support of the Microelectronic System Laboratory of Ecole Polytechnique Fédérale de Lausanne, and to thank the Beckman Institute for Advanced Science and Technology for its hospitality during his research.

**Appendix.**

**Appendix A.**

Eq.3 is of the general form

$$\sum_{\substack{\mathbf{k}_1,\mathbf{k}_{1'}\\ \mathbf{k}_2,\mathbf{k}_{2'}}} H(\mathbf{k}_1,\mathbf{k}_{1'};\mathbf{k}_2,\mathbf{k}_{2'})\delta(k_1-k_2-k_{1'}-k_{2'})\delta_{\mathbf{k}_1+\mathbf{k}_2,\mathbf{k}_{1'}+\mathbf{k}_{2'}} \quad (A.1)$$

where $H$ is a general function of the wavevectors $\mathbf{k}_1,\mathbf{k}_{1'},\mathbf{k}_2,\mathbf{k}_{2'}$ (Fig. A.1). To evaluate the 8-uple summations, we first exploit one identity of the delta function,

$$\delta\left[(k_1-k_2)^2-(k_{1'}+k_{2'})^2\right]=\frac{\delta(k_1-k_2-k_{1'}-k_{2'})+\delta(k_1-k_2+k_{1'}+k_{2'})}{2(k_1-k_2)} \quad (A.2)$$

where the second term on the right hand side (RHS) is zero since the available phase-space is restricted by energy conservation. We then eliminate the summation over $\mathbf{k}_{2'}$ by using the momentum-conservation Kronecker delta,

$$\sum_{\mathbf{k}_1,\mathbf{k}_{1'},\mathbf{k}_2} 2H(\mathbf{k}_1,\mathbf{k}_{1'};\mathbf{k}_2,\mathbf{k}_{2'})(k_1-k_2)\delta\left[(k_1-k_2)^2-k_{1'}^2-2k_1k_2-(\mathbf{k}_1+\mathbf{k}_2-\mathbf{k}_{1'})^2\right] \quad (A.3)$$

where $\mathbf{k}_{2'}=\mathbf{k}_1+\mathbf{k}_2-\mathbf{k}_{1'}$. After expanding the last term in the argument of the $\delta$-function and using momentum conservation, we get

$$\sum_{\mathbf{k}_1,\mathbf{k}_{1'},\mathbf{k}_2} H(\mathbf{k}_1,\mathbf{k}_{1'};\mathbf{k}_2,\mathbf{k}_{2'})\frac{(k_1-k_2)}{2}\delta\left\{k_1k_2\cos^2\left[(\theta_1-\theta_2)/2\right]+k_1k_{2'}\sin^2\left[(\theta_{1'}-\theta_{2'})/2\right]\right\} \quad (A.4)$$

where $\theta_1,\theta_2,\theta_{1'},\theta_{2'}$ are the angles of the respective wavevectors with respect to the field direction. Now, we transform the sums over the wavevectors $\mathbf{k}_{1'},\mathbf{k}_2$ into integrals over their respective magnitudes and angles.

$$\frac{A^2}{(2\pi)^4}\sum_{\mathbf{k}_1}\iint k_1 k_2 dk_1 dk_2 \int_0^{2\pi} d\theta_{1'} \int_0^{2\pi} d\theta_2 H(\mathbf{k}_1,\mathbf{k}_{1'};\mathbf{k}_2,\mathbf{k}_{2'})\frac{(k_1-k_2)}{2}\times$$
$$\times \delta\{k_1 k_2 \cos^2[(\theta_1-\theta_2)/2] + k_{1'}k_{2'}\sin^2[(\theta_{1'}-\theta_{2'})/2]\} \tag{A.5}$$

Let us focus on the double angular integral.

$$\int_0^{2\pi} d\theta_{1'} \int_0^{2\pi} d\theta_2 H(\theta_{1'},\theta_2)\delta\{k_1 k_2 \cos^2[(\theta_1-\theta_2)/2] + k_{1'}k_{2'}\sin^2[(\theta_{1'}-\theta_{2'})/2]\} \tag{A.6}$$

Here, $H$ is the same function as in eq. A.1 where the dependence on other variables is omitted for brevity. The $\delta$-function is of the form $\delta[\alpha^2+\beta^2]$ with

$$\alpha \equiv \sqrt{k_1 k_2}\cos[(\theta_1-\theta_2)/2], \beta \equiv \sqrt{k_{1'}k_{2'}}\sin[(\theta_{1'}-\theta_{2'})/2] \tag{A.7}$$

By changing the variables $\theta_2,\theta_{1'} \to \alpha,\beta$, the double angular integral can be transformed into a simpler form,

$$\int_{-\sqrt{k_1 k_2}}^{\sqrt{k_1 k_2}} \frac{d\alpha}{|\partial\alpha/\partial\theta_2|} \int_{-\sqrt{k_{1'}k_{2'}}}^{\sqrt{k_{1'}k_{2'}}} \frac{d\beta}{|\partial\beta/\partial\theta_{1'}|} \cdot H[\theta_2(\alpha),\theta_{1'}(\beta)]\delta[\alpha^2+\beta^2] \tag{A.8}$$

One notes that since the only contribution to the integral occurs at the origin, we can extend the integral limits to the entire $\alpha, \beta$ plane,

$$\int_{-\infty}^{\infty} \frac{d\alpha}{|\partial\alpha/\partial\theta_2|} \int_{-\infty}^{\infty} \frac{d\beta}{|\partial\beta/\partial\theta_{1'}|} \cdot H[\theta_2(\alpha),\theta_{1'}(\beta)]\delta[\alpha^2+\beta^2] \tag{A.9}$$

and proceed to changing the Cartesian coordinates $\alpha, \beta$ into the polar coordinates $\rho, \phi$. This yields

$$\int_0^{2\pi} \frac{d\varphi}{|\partial\alpha/\partial\theta_2|} \int_0^\infty \frac{\rho d\rho}{|\partial\beta/\partial\theta_{1'}|} \cdot H\{\theta_2[\alpha(\rho,\varphi)], \theta_{1'}[\beta(\rho,\varphi)]\} \delta(\rho^2)$$

$$= \frac{1}{2} \int_0^{2\pi} \frac{d\varphi}{|\partial\alpha/\partial\theta_2|} \int_0^\infty \frac{d\rho}{|\partial\beta/\partial\theta_{1'}|} \cdot H\{\theta_2[\alpha(\rho,\varphi)], \theta_{1'}[\beta(\rho,\varphi)]\} \delta(\rho) \quad (A.10)$$

where we have used the identity $2\rho\delta(\rho^2) = \delta(\rho)$. After carrying out the integrals, we get

$$\frac{\pi}{2|\partial\alpha/\partial\theta_2|_{\alpha=0} |\partial\beta/\partial\theta_{1'}|_{\beta=0}} H[\theta_2(\alpha=0), \theta_{1'}(\beta=0)]$$

$$= \frac{2\pi}{\sqrt{k_1 k_2 k_{1'} k_{2'}} \sin[(\theta_1-\theta_2)/2]_{\theta_2=\theta_1-\pi} \cos[(\theta_{1'}-\theta_{2'})/2]_{\theta_{1'}=\theta_{2'}}} H(\theta_2=\theta_1-\pi, \theta_{1'}=\theta_{2'}) \quad (A.11)$$

It should be noted that $\alpha=0, \beta=0$ is equivalent to $\theta_2=\theta_1-\pi, \theta_{1'}=\theta_{2'}$, respectively. This allows us to write the original expression (A.1) as

$$\sum_{\substack{\mathbf{k}_1,\mathbf{k}_{1'} \\ \mathbf{k}_2,\mathbf{k}_{2'}}} H(\mathbf{k}_1,\mathbf{k}_{1'};\mathbf{k}_2,\mathbf{k}_{2'}) \delta(k_1-k_2-k_{1'}-k_{2'}) \delta_{\mathbf{k}_1+\mathbf{k}_2,\mathbf{k}_{1'}+\mathbf{k}_{2'}}$$

$$= \frac{A^2}{16\pi^3} \sum_{\mathbf{k}_1} \frac{\iint k_1 k_2 dk_1 dk_2 \int_0^{2\pi} d\theta_{1'} \int_0^{2\pi} d\theta_2 H(\mathbf{k}_1,\mathbf{k}_{1'};\mathbf{k}_2,\mathbf{k}_{2'}) \times}{\times \frac{(k_1-k_2)}{\sqrt{k_1 k_2 k_{1'} k_{2'}}} \frac{\delta(\theta_2-\theta_1+\pi)\delta(\theta_{1'}-\theta_{2'})}{\sin[(\theta_1-\theta_2)/2]\cos[(\theta_{1'}-\theta_{2'})/2]}} \quad (A.12)$$

which reduces our six-fold summation over $\mathbf{k}_{1'}, \mathbf{k}_2, \mathbf{k}_{2'}$ into a two-dimensional integral over $k_2, k_{1'}$, and finally write

$$= \frac{A^2}{16\pi^3} \sum_{\mathbf{k}_1} \int_0^\infty dk_{1'} k_{1'} \int_0^\infty dk_2 k_2 \frac{(k_1-k_2)}{\sqrt{k_1 k_2 k_{1'} k_{2'}}} H(\mathbf{k}_1 \mathbf{k}_2; \mathbf{k}_{1'} \mathbf{k}_{2'})|_{\substack{\mathbf{k}_{2'}=\mathbf{k}_1+\mathbf{k}_2-\mathbf{k}_{1'}; \\ \theta_1=\theta_2+\pi; \theta_{1'}=\theta_{2'}}} \quad (A.13)$$

where $\theta_{2'}$ represents the angle of the vector $\mathbf{k}_1+\mathbf{k}_2-\mathbf{k}_{1'}$ all being co-linear.

The expression (A.13) is equivalent to eq.3 in the original publication [11] and its erratum [19] except for a factor ½, which is readily corrected.

**Appendix B.**
Eq.11 is obtained as follows:

After multiplying both sides of eq.10 by $1+F/F_C$ and integrating from source to the x-position in the channel [15], one gets

$$(1+\frac{V(x)}{F_C})\frac{I}{W} = C_{ox}\mu_0[V_{GT}-V(x)/2]V(x) + 2e\int_0^x(1+\frac{F}{F_C})dx'\int_0^{x''}G(x'')dx''. \quad (B.1)$$

The double integral in the last term of eq.B.1 can be integrated by parts, yielding

$$\int_0^x(1+\frac{F}{F_C})dx'\int_0^{x''}G(x'')dx'' = \int_0^{x'}G(x'')dx''\int_0^{x'}(1+\frac{F}{F_C})dx''\bigg|_0^x - \int_0^xG(x')dx'\int_0^{x'}(1+\frac{F}{F_C})dx''$$
$$= (x+\frac{V(x)}{F_C})\int_0^xG(x')dx' - \int_0^xG(x')(x'+\frac{V(x')}{F_C})dx' \quad (B.2)$$

After taking the limit $x \rightarrow L$ and rearranging eqs.(B.1 & 2) one obtains

$$I = \frac{C_{ox}W\mu_0[V_{GT}V_{DS}-V_{DS}^2/2]}{L(1+\frac{V_{DS}}{V_C})} + \frac{2eW}{1+\frac{|V_{DS}|}{V_C}}\int_0^L\left(1-\frac{x'}{L}+\frac{|V_{DS}|-|V(x)|}{V_C}\right)G(x')dx' \quad (B.3)$$

where I is the total current in the channel and $V_C = F_C L$.

**Appendix C.**
Eq.12 for the electric field along the channel is obtained by combining eqs.(B.1 & 2), which yields

$$(1+\frac{V}{F_C})\frac{I}{W} = C_{ox}\mu_0[V_{GT}-V/2]V + 2e(x+\frac{V}{F_C})\int_0^xG(x')dx' - 2e\int_0^xG(x')(x'+\frac{V(x')}{F_C})dx' \quad (C.1)$$

from which one obtains an implicit expression for the potential along the channel by solving the second order algebraic equation in $V$

$$V(x) = \left\{ \frac{V_{GT} - V_i + \frac{2e}{F_C C_{ox} \mu_0} \int_0^x G(x')dx' -}{\sqrt{\left[V_{GT} - V_i + \frac{2e}{F_C C_{ox} \mu_0} \int_0^x G(x')dx'\right]^2 - 2F_C V_i x + \frac{4e}{C_{ox} \mu_0} \int_0^x \left[x - x' + \frac{V(x)}{F_C}\right] G(x')dx'}} \right\} \quad (C.2)$$

where $V_i = \frac{I}{W \mu_0 F_C C_{ox}}$.

Deriving eq.C.2 with respect to position $x$ yields the electric field

$$F(x) = -\left\{ \frac{\frac{2eG(x)}{F_C C_{ox} \mu_0} - \frac{\frac{2eG(x)}{F_C C_{ox} \mu_0}\left[V_{GT} - V_i - V(x) + \frac{2e}{F_C C_{ox} \mu_0}\int_0^x G(x')dx'\right] + \frac{2e}{C_{ox}\mu_0}\int_0^x G(x')dx' - F_C V_i}{\sqrt{\left[V_{GT} - V_i + \frac{2e}{F_C C_{ox} \mu_0} \int_0^x G(x')dx'\right]^2 - 2F_C V_i x + \frac{4e}{C_{ox}\mu_0} \int_0^x \left[x - x' + \frac{V(x)}{F_C}\right] G(x')dx'}}}{} \right\}.$$

(C.3)

**References**

[1] A.K. Geim and K. S. Nonselov, *Nat. Mater.*, vol. **6**, pp. 183-191 (2007)

[2] F. Xia, D. B. Farmer, Y. Lin and P. Avouris, *Nano Letters*, vol. **10**, no. 2, pp.715-718 (2010)

[3] M. Wilson, *Phys. Today*, vol. **59**, no. 1, pp. 21 (2006)

[4] Y. Zhang,Y. W. Tan, H. L. Stormer, and P. Kim, *Nature*, vol. **438**, 201 (2005)

[5] C. Berger, Z. Song, X. Li, X. Wu, N. Brown, C. Naud, D. Mayou, T. Li, J. Hass, A. N. Marchenkov, E. H. Conrad, P. N. First, and W. A. de Heer, *Science*, vol. **312**, 1191 (2006)

[6] A. Akturk and N. Goldsman, *J. Appl. Phys.*, vol. **103**, 053702 (2008)

[7] X. Ling, L. Xie, Y. Fang, H. Xu, H. Zhang, J. Kong, M. S. Dresselhaus, J. Zhang and Z. Liu, *Nano Letters*, vol. **10**, no. 2, pp. 553-561 (2010)

[8] B. G. Streetman and S. K. Banerjee, *Solid State Electronic Devices,* Sixth Ed., New Jersey: Pearson Prentice Hall. 2006

[9] X. Li, E. A. Barry, J. M. Zavada, M. Buongiorno Nardelli, and K. W. Kim, *Appl. Phys. Lett*. vol. **97**, 082101 (2010)

[10] C. L. Anderson and C. R. Crowell, *Physical Review B*, vol. **5**, no. 6, pp. 2267-2272 (1972)

[11] A. Girdhar and JP. Leburton, *Appl. Phys. Lett*., vol. **99**, 043107 (2011)

[12] I. Meric, M. Y. Han, A. F. Young, B. Ozyilmaz, P. Kim and K. L. Shepard, *Nature Nanotechnol.*, vol. **3**, pp.654-659 (2008)

[13] G. Baurer, *Determination of Electron Temperatures and of Hot Electron Distribution Functions in Semiconductros,* Berlin: Springer. 1974

[14] K. Hess, *Advanced Theory of Semiconductor Devices*, New York: Wiley-IEEE Press. 2000

[15] R. S. Muller and T. I. Kamis, *Devices Electronics for Integrated Circuits,* 3$^{rd}$ ed. New York: Wiley, 2003

[16] K. Hess and P. Vogl, *Solid St. Comm.*, vol. **30**, pp. 807 (1979)

[17] W. Shockley, *BellSystem Tech. J.*, vol. **30**, pp. 990 (1951)

[18] B. Scott and JP. Leburton, *IEEE Trans. Nanotachnol.*, vol. **10**, no. 5 pp. 1113-1119, 1(2011)

[19] A. Girdhar and JP. Leburton, *Appl. Phys. Lett*., vol. **99**, 229903 (2011)

**Figure Captions**

FIG. 1. Schematic of a carrier multiplication scattering event in the Dirac cone in graphene. Conduction electron **1** collides with valence electron **2** to occupy new states **1'** and **2'** in the conduction band and create a valence hole. The reverse event is Auger recombination.

FIG. 2. Current densities as a function of applied fields for temperature 300-1200 K at various carrier concentrations. Squares are values obtained from ref. [11] and the solid lines are the data best fit. The carrier concentrations for each electronic temperature are $10^{12}$cm$^{-2}$, $2\times10^{12}$cm$^{-2}$, $5\times10^{12}$cm$^{-2}$, $10^{13}$cm$^{-2}$ (from top to down).

FIG. 3. a) $\alpha$-coefficient as a function of carrier concentration. b) Net generation rate as a function of carrier concentration for several values of electronic temperature. c) $n_0$-coefficient as a function of electronic temperature.

FIG. 4. Schematic of G-FET device of channel length $L$ (source-drain separation).

FIG. 5. Electric field as a function of position in the channel for different values of drain source voltage before saturation.

Fig. 6. Flow-chart of the interaction scheme to calculate the current by taking into account the generation rate self-consistently.

FIG. 7. a) Drain current as a function of drain source voltage for several values of $V_{GT}$:1V, 1.5V and 2V (from bottom to up). b) Electric field as a function of drain source voltage for several values of $V_{GT}$:1V, 1.5V and 2V (from top to down).

FIG. 8. a) Net generation rate and electronic temperature as a function of position for different values of $V_{DS}$=1.5V, 2V and 2.5V (from bottom to up). b) Net generation rate as a function of drain source voltage for different values of critical field: 25kV/cm, 35kV/cm, 45kV/cm and 55kV/cm (from top to down).

FIG. 9. a) Drain current as a function of source drain voltage for $V_{gback}$=-40V and several values of $V_{gtop}$:0V, -0.3V and -0.8V (from bottom to up) and different critical field:16kV/cm for 0V and 22kV/cm for the other two curves. b) Drain current as a function of source drain voltage for $V_{gback}$=40V and several values of $V_{gtop}$:-0.3V, -0.8V and -1.3V (from bottom to up) and different critical field:15kV/cm for the first two curves and 22kV/cm for the highest. The dashed curve is got with $R_S$=600Ω and $F_{CT}$=18kV/cm.

FIG. 10. a) Carrier concentration as a function of source drain voltage for $V_{gback}$=40V and several values of $V_{gtop}$:-0.3V, -0.8V, -1.3V (from bottom to up). b) Carrier Velocity as a function of normalized source drain voltage respect to saturation for $V_{gback}$=40V and several values of $V_{gtop}$:-0.3V, -0.8V and -1.3V (from bottom to up).

FIG. A. 1. The wavevectors $k_1, k_{1'}, k_2, k_{2'}$ and their angles measured with respect to the electric field direction.

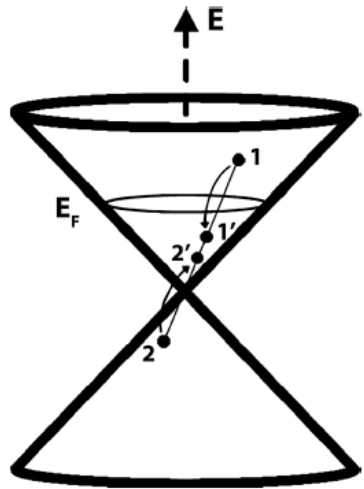

Figure 1

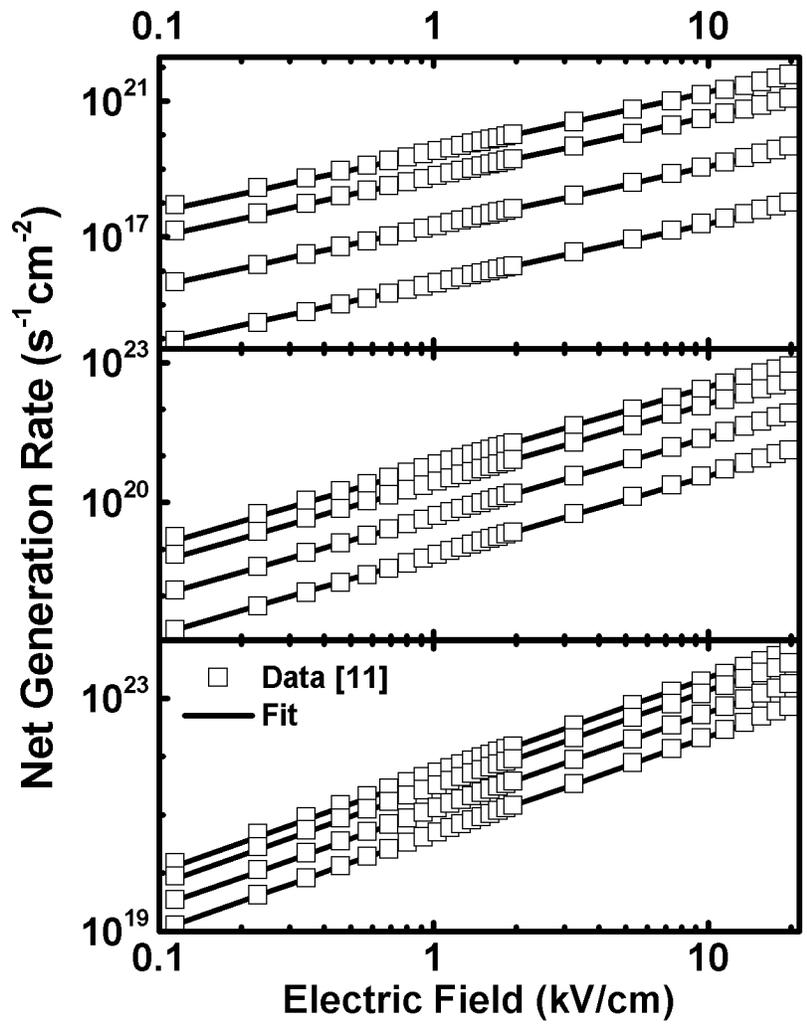

Figure 2

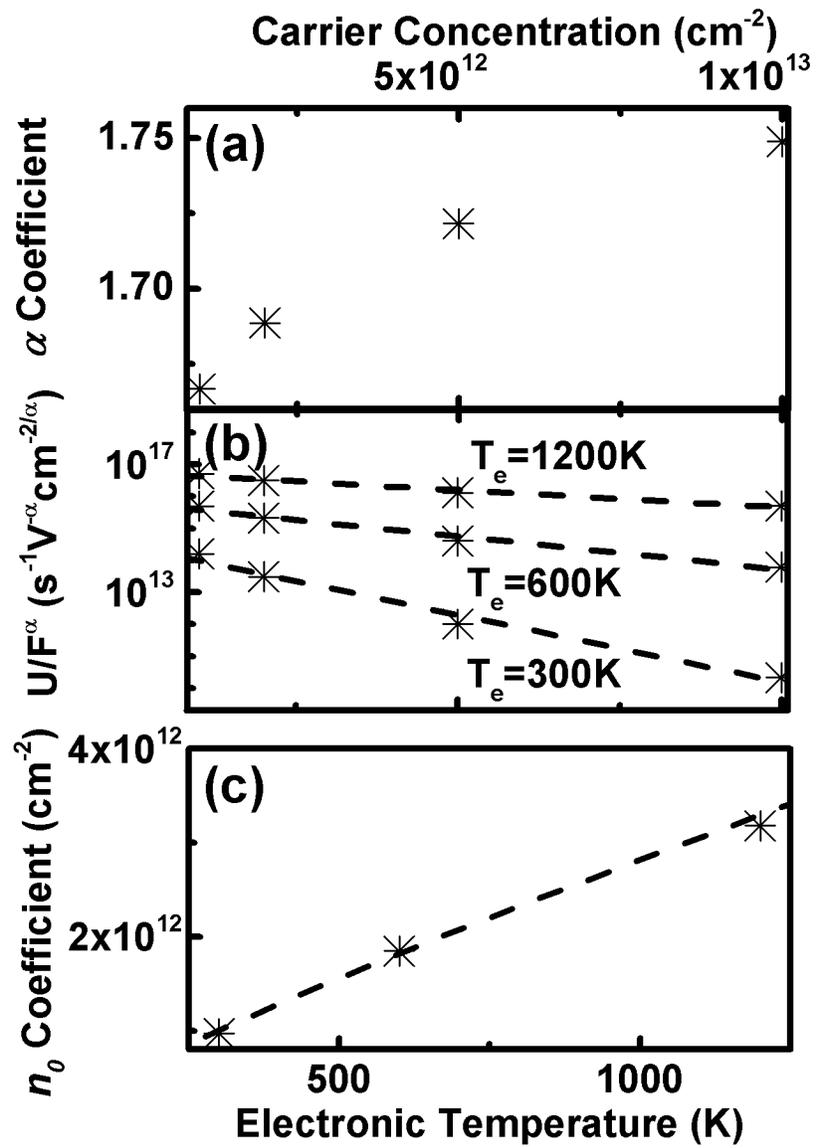

Figure 3

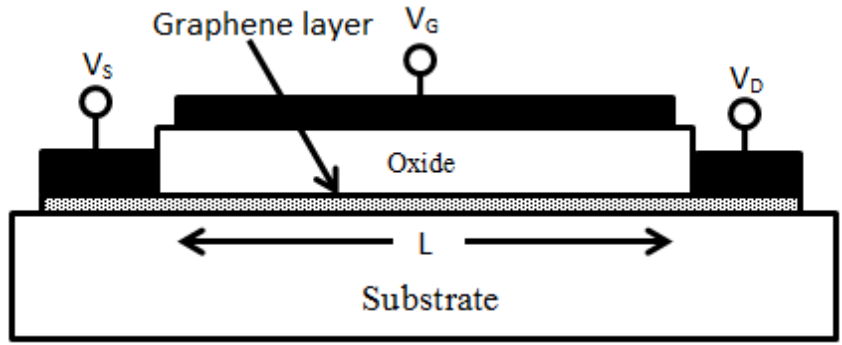

Figure 4

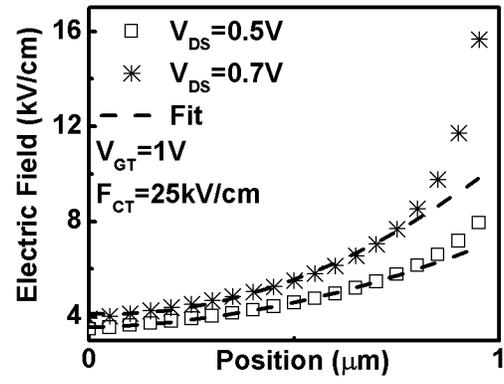

Figure 5

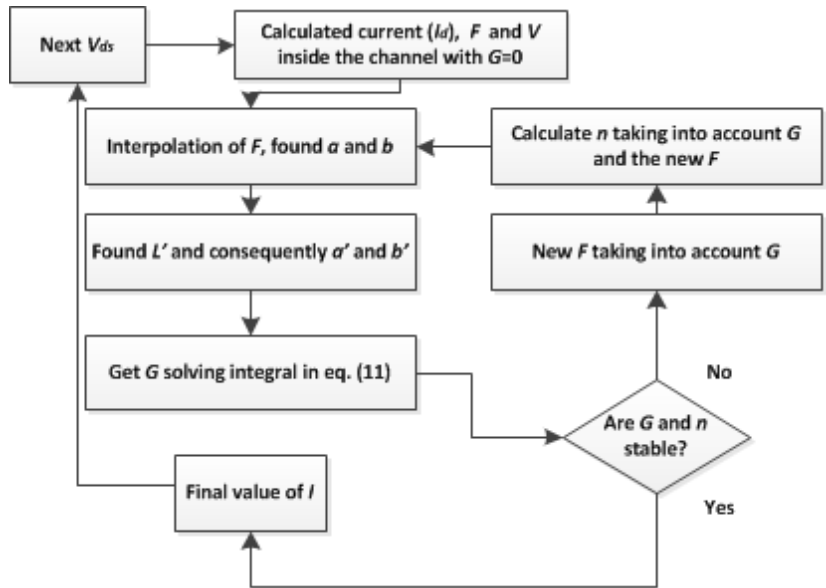

Figure 6

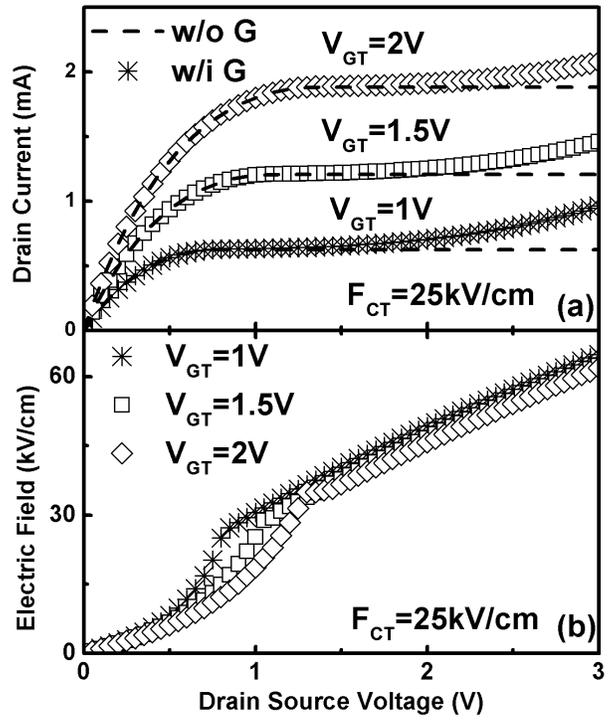

Figure 7

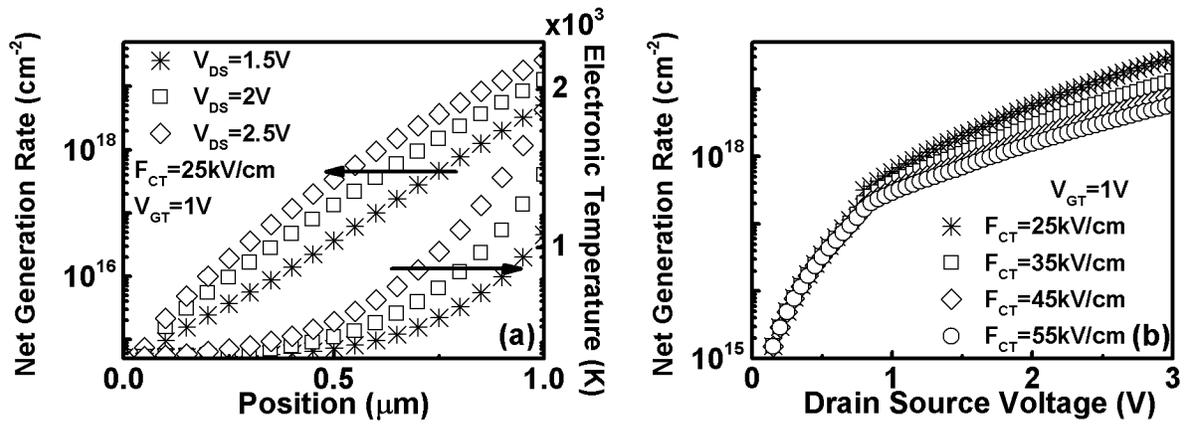

Figure 8

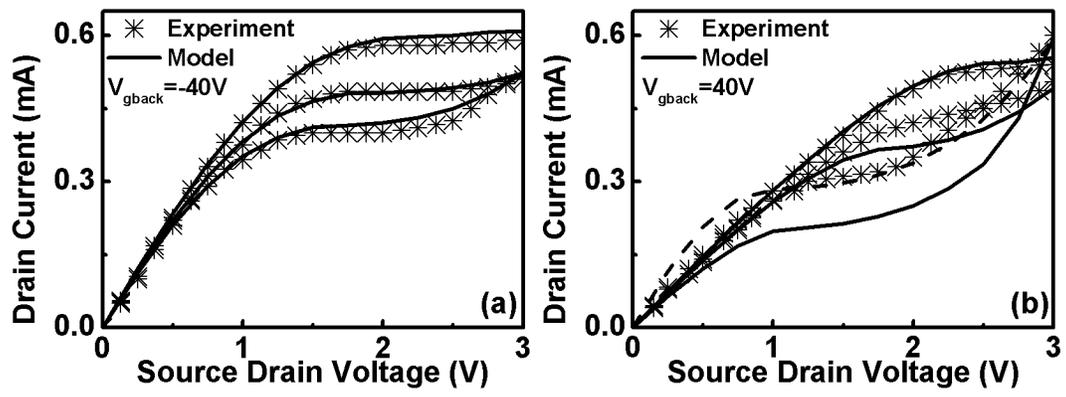

Figure 9

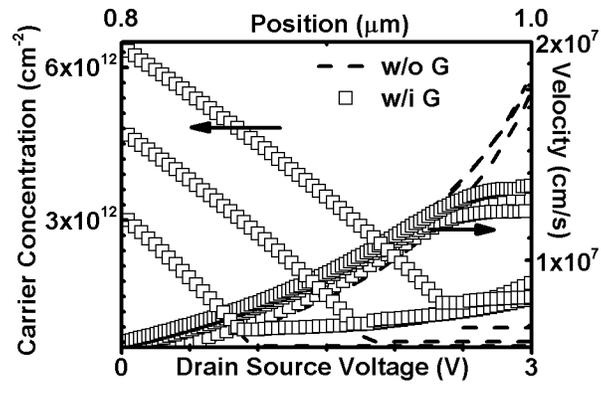

Figure 10

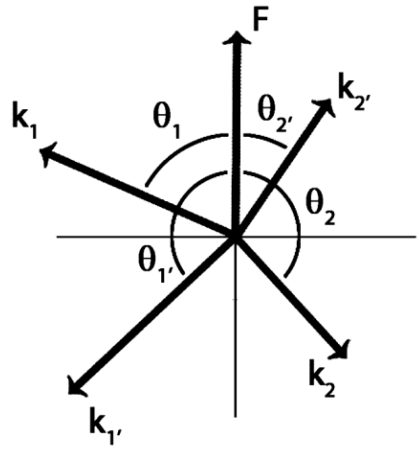

Figure A.1